\documentclass[12pt]{article}
\usepackage{graphicx}

\newcommand{\mnui}{\mbox{$m(\nu_{\mathrm i})$}}

\newcommand{\mnue}{\mbox{$m(\nu_{\mathrm e} )$}}

\newcommand{\mtwonue}{\mbox{$m^2(\nu_{\mathrm e} )$}}

\newcommand{\mee}{\mbox{$m_\mathrm{ee}$}}
\newcommand{\mtwo}{\mbox{$m_\nu^2$}}

\newcommand{\ttwo}{\mbox{$\rm T_2$}}

\newcommand{\kr}{\mbox{$\rm ^{83m}Kr$}}
\newcommand{\rhenium}{\mbox{$\rm ^{187}Re$}}

\newcommand{\ev}{\mbox{$\rm eV/c^2$}}

\newcommand{\bdec}{\mbox{$\beta$~decay}}
\newcommand{\bspec}{\mbox{$\beta$~spectrum}}
\newcommand{\belec}{\mbox{$\beta$~electron}}

\newcommand{\ezero}{\mbox{$E_0$}}

\newcommand{\be}{\begin{equation}}
\newcommand{\ee}{\end{equation}}
\newcommand{\bea}{\begin{eqnarray}}
\newcommand{\eea}{\end{eqnarray}}

\newcommand{\etal}{\mbox{\it et al.}}

\title{Non-oscillation probes of neutrino masses}

\author{C. Weinheimer\\
\it \footnotesize
Institut f\"ur Kernphysik, Westf\"alische
Wilhelms-Universit\"at M\"unster, 48149 M\"unster, Germany,\\
\it \footnotesize Email: weinheimer@uni-muenster.de
}

\begin{document}
\maketitle

\begin{abstract}
The absolute scale of neutrino masses is very important for understanding the evolution and the structure
formation of the universe as well as for nuclear and particle physics beyond the present Standard Model.
Complementary to deducing statements on the neutrino mass from cosmological observations two different methods
to determine the neutrino mass scale in the laboratory are pursued: the search for neutrinoless double \bdec\ and the 
direct neutrino mass search. For both methods currently experiments with a sensitivity of {\cal O}(100)~meV are being 
set up or commissioned.

\end{abstract}

\section{Introduction}

Neutrino oscillation experiments have shown recently, that the
different neutrino flavors mix and can oscillate during flight from one 
flavor state into another. The analysis of all neutrino oscillation 
experiments yields the mixing angles 
and the differences of squared neutrino mass eigenstates  
\cite{nu_osc}.
Clearly these findings prove that neutrinos have non-zero masses, but they 
cannot determine the absolute neutrino mass scale. 
The huge abundance of neutrinos left over in the universe from the big bang 
($\approx 336$/cm$^3$) and their contribution to structure formation 
and the evolution of the universe ({\it e.g.}\cite{hannestad})
as well as the key role of neutrino masses in finding the new Standard Model 
of particle physics ({\it e.g.} \cite{nu_mass_theo}) 
make the absolute value of the neutrino mass one of todays 
most urgent questions of astroparticle physics
and cosmology as well as of nuclear and particle physics.

\begin{figure}[t!]
\centerline{\includegraphics[width=0.8\textwidth]{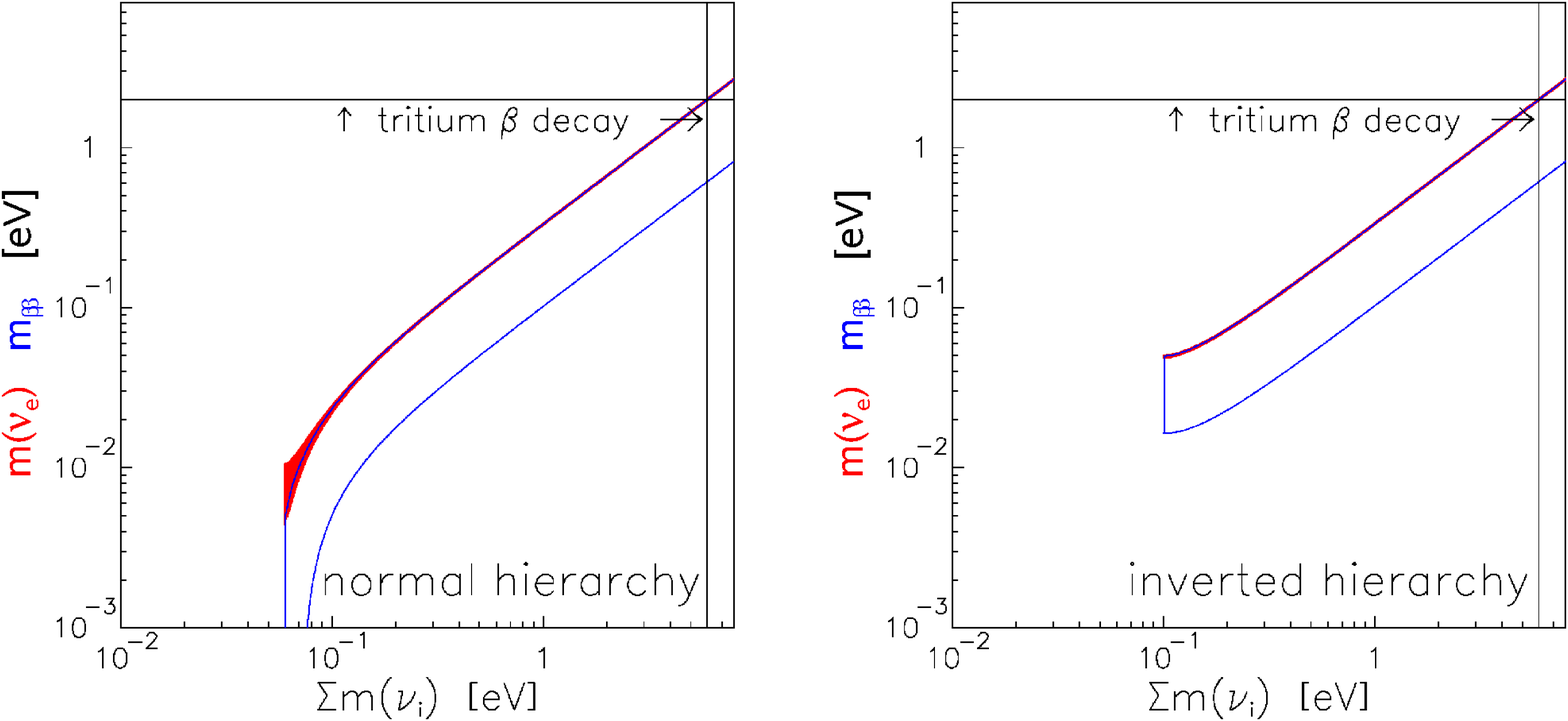}}    
\caption{Observables of neutrinoless double \bdec\ \mee\ (open band) and of direct neutrino mass determination by single \bdec\ \mnue\ 
(gray band on upper boarder of \mee\ band) versus the cosmologically relevant sum of neutrino mass eigenvalues 
$\sum \mnui$ for the case of normal hierarchy (left) and of inverted hierarchy (right) \cite{otten08}. 
The width of the bands/areas is caused by the experimental uncertainties of the neutrino mixing angles \cite{pdg08}
and in the case of \mee\ also by the completely unknown Majorana- and CP-phases. 
Uncertainties of the nuclear matrix elements, which enter \mee , are not considered.
\label{fig:comparison_methods}} 
\end{figure}

There exist 3 different approaches to the absolute neutrino mass scale: 
\begin{itemize}
\item Cosmology\\
  Essentially the size of fluctuations is observed at different scales by using cosmic microwave background and
  large scale structure data. Since the light neutrinos would have smeared out fluctuations at small scales the power spectrum
  at small scales is sensitive to the neutrino mass. Up to now, only limits on the sum of the 3 neutrino masses have been obtained around
  $\sum \mnui < 0.61~\ev$ \cite{wmap2008}, which are to some extent 
model- and analysis dependent \cite{hannestad_bao07}.
\item Neutrinoless double \bdec\ ($0\nu\beta\beta$)\\
  A neutrinoless double {\bdec} (two \bdec s in the same nucleus at the same time with emission of two \belec s (positrons) while the (anti)neutrino
  emitted at one vertex is absorbed at the other vertex as a neutrino (antineutrino)) 
  is forbidden in the Standard Model of particle physics. It could exist, if the neutrino is its own antiparticle (``Majorana-neutrino'' in
  contrast to ``Dirac-neutrino''). Furthermore, a finite neutrino mass is 
  required in order to produce in the chirality-selective interaction a neutrino with a small component of opposite handedness on 
  which this neutrino exchange subsists. The decay rate would scale with the absolute square of the so called effective neutrino mass,
  which takes into account the neutrino mixing matrix $U$:
  \be\
    \Gamma_{0\nu\beta\beta} \propto \left| \sum U^2_{\rm ei} \mnui \right|^2 := \mee^2
  \ee
Here \mee\ represents the coherent sum of the \mnui-components of the $0\nu\beta\beta$-decay amplitudes and hence carries their relative phases
  (the usual CP-violating phase of an unitary $3 \times 3$ mixing matrix plus two so-called Majorana-phases). A significant additional uncertainty
  entering the relation of \mee\ and the decay rate is the nuclear matrix element of the neutrinoless double \bdec . 
\item Direct neutrino mass determination\\
  The direct neutrino mass determination is based purely on relativistic kinematics without further assumptions. 
  Therefore it is sensitive to the neutrino mass squared $m^2(\nu)$.
  In principle there are two methods: time-of-flight measurements and precision investigations of weak decays.
  The former requires very long baselines and therefore
  very strong sources, which only cataclysmic cosmological events like a core-collapse supernova could
  provide.  The non-observation of a dependence of the arrival time on energy of supernova neutrinos from SN1987a gave 
  an upper limit on the neutrino mass of 5.7~\ev\ \cite{pdg08}. 
  Unfortunately nearby supernova explosions are too rare
  and too little understood to allow an improvement into the sub-eV range.

  Therefore, aiming for this sensitivity, the investigation of the kinematics of weak decays and more explicitly the
  investigation of the endpoint region of a \bdec\ spectrum is still the most sensitive model-independent and direct method 
  to determine the neutrino mass. 
  Here the neutrino is not observed but the charged decay products are precisely measured. Using energy and momentum conservation 
  the neutrino mass can be obtained. In the case of the investigation of a \bspec\ usually the ``average electron neutrino mass''
  \mnue\ is determined:
  \be\ \label{eq_define_mnue}
    \mnue^2 := \sum |U^2_{\rm ei}| \mnui^2
  \ee\ 
   This incoherent sum is not sensitive to phases of the neutrino mixing matrix.
\end{itemize}  
Figure \ref{fig:comparison_methods} demonstrates that the different methods are complementary to each other and 
compares them. It shows, that the cosmological relevant neutrino mass scale  $\sum \mnui$ has a nearly 
one-to-one correlation to \mnue\ determined by direct neutrino mass experiments. The observable of 
neutrinoless double \bdec\ does not allow a very precise neutrino mass determination due to the
unknown Majorana phases and the uncertainties of the nuclear matrix elements. On the other hand the combination
of all three methods gives an experimental handle on the Majorana phases. Furthermore, the search for the neutrinoless
double \bdec\ is the only way to prove the Majorana character of neutrinos and one of the most promising ways to search for lepton number violation. 

This article is structures as follows: Section 2 reports on the various searches for neutrinoless double \bdec . 
In section 3 the neutrino mass determination from tritium and \rhenium\ \bdec\ are presented.
The conclusions are given in section 4.

\section{Search for neutrinoless double $\bf \beta$ decay}
There are more than 10 double \bdec\ isotopes. For most of them the normal double \bdec\ with neutrino emission has been observed. 
For neutrinoless double \bdec\ there is only one claim for evidence at $\mee \approx 0.4 ~\ev$ by part of the Heidelberg-Moscow 
collaboration \cite{klapdor04}, all other experiments so far set upper limits.

The signature of neutrinoless double \bdec\ is, that the sum of the energy of both decay electrons 
(in case of double $\beta^-$ decay,  positrons for double $\beta^+$ decay) is equal to the $Q$-value of the nuclear transition. 
The experimental approaches can be classified into two methods (see figure \ref{fig:dbd_methods}):

\begin{figure}[t!]
\centerline{\includegraphics[width=0.8\textwidth]{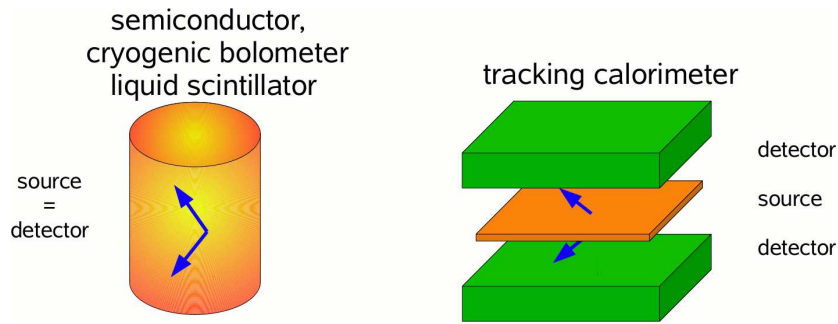}}
\caption{Two different experimental configurations in search for the neutrinoless double \bdec .
\label{fig:dbd_methods}} 
\end{figure}

\subsection{``Source$=$detector'' configuration}
In the ``source$ = $detector'' configuration the double \bdec\ nuclei are part of the detector, 
  which measures the sum of the energy of both \belec s. 
  The experimental implementation of these calorimeters are semiconductors ({\it e.g.} isotope: $^{76}$Ge, experiments: GERDA, MAJORANA),
  cryo-bolometers ({\it e.g.} isotope: $^{130}$Te, experiments: CUORICINO and its successor CUORE) and liquid scintillators ({\it e.g.} isotope:
  $^{136}Ge$, $^{152}$Nd, experiments: EXO-200, SNO+).
  In general, this method allows more easily to install a large target mass.
  Currently the most sensitive limits comes from the CUORICINO experiment 
  at the Gran Sasso underground laboratory LNGS 
  yielding $t_{1/2}(^{130}$Te$) > 3 \cdot 10^{24}$~y and 
  $m_{\rm ee} < 0.19 - 0.68$~eV \footnote{The range of the effective 
  neutrino mass limits originates from the different nuclear matrix elements
  calculated by different theory groups.} \cite{cuoricino08}.
  The CUORICINO experiment has been completed and the installation of the 
  successor experiment CUORE has started.   
  The installation of the GERDA experiment is nearly finished and the 
  commissioning will start end of 2009.

\subsection{``Source$\neq$detector'' configuration}
In the this configuration the double \bdec\ source is separated from two tracking calorimeters, which determine direction
 and energy of both \belec s separately ({\it e.g.} isotope $^{82}$Se, $^{100}$Mo, experiments: NEMO3 and its successor SuperNEMO, ELEGANT).
  By this method the most sensitive limit comes from the NEMO3 experiment
  in the Modane underground laboratory LSM yielding  $t_{1/2}(^{100}$Mo$) > 1.1 \cdot 10^{24}$~y and 
  $m_{\rm ee} < 0.45 - 0.93$~eV $^1$ \cite{mauger_taup09}.  
 Although it requires much larger detectors to accumulate similar large target masses as in the ``source$=$detector'' case, there is the advantage,
  that the independent information of both electrons allows to study double \bdec\ processes with 2 neutrinos in detail. In case neutrinoless
  double \bdec\ would be detected, the angular correlation of both electrons will allow to draw some conclusion on the underlying process 
  \footnote{A theorem by Schechter and Valle \cite{schechtervalle} requests the neutrinos to have non-zero Majorana masses, 
  if neutrinoless double \bdec\ is proven to exist, but the dominant process could still be different, {\it e.g.} based on the exchange
  of a SUSY particle or by the contribution of right-handed weak charged currents.}.

\section{Direct neutrino mass experiments}

\begin{figure}[t!]
\centerline{\includegraphics[width=0.7\textwidth]{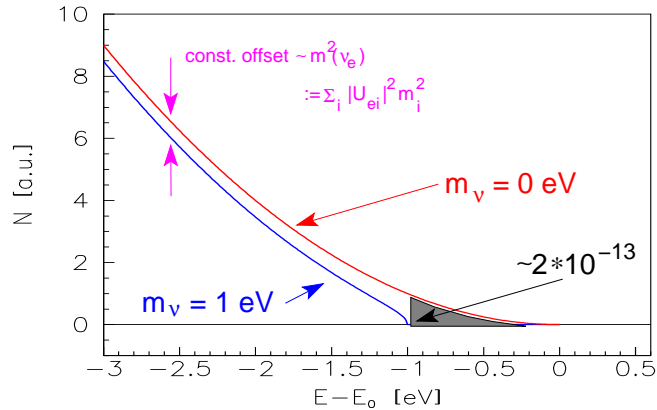}}
\caption{Expanded \bspec\ around its endpoint \ezero\
   for $\mnue = 0$ (dashed line) and for an arbitrarily chosen neutrino mass
  of 1~eV (solid line).
  In the case of tritium, the gray-shaded area corresponds to a fraction of $2 \cdot 10^{-13}$ of all
  tritium \bdec s.
\label{fig:beta_spec}}
\end{figure}

The signature of a non-zero neutrino mass is a tiny 
modification of the spectrum of the \belec s near its endpoint (see figure \ref{fig:beta_spec}),
which has to be measured with very high precision. To maximize this effect,
$\beta$ emitters with low endpoint energy are favored.

\subsection{``Source$\neq$detector'' configuration: Tritium $\beta$ decay experiments}
 
Tritium is the standard isotope for this kind of study due to 
its low endpoint of 18.6~keV, its rather short half-life of 12.3~y,
its super-allowed shape of the \bspec ,
and its simple electronic structure. 
Tritium \bdec\ experiments using a tritium source and a separated 
electron spectro\-meter have been performed in search for the neutrino mass for more than 50~years \cite{otten08}
yielding  a sensitivity of 2~eV by the experiments at Mainz \cite{kraus05} and Troitsk \cite{lobashev03}.
The huge improvement of these experiments in the final sensitivity as well as
in solving the former ``negative \mtwonue `` problem with
respect to previous experiments is mainly caused by the new spectrometers
of MAC-E Filter type and by careful studies of the systematics.

\begin{figure}
\centerline{\includegraphics[angle=0,width=0.7\textwidth]{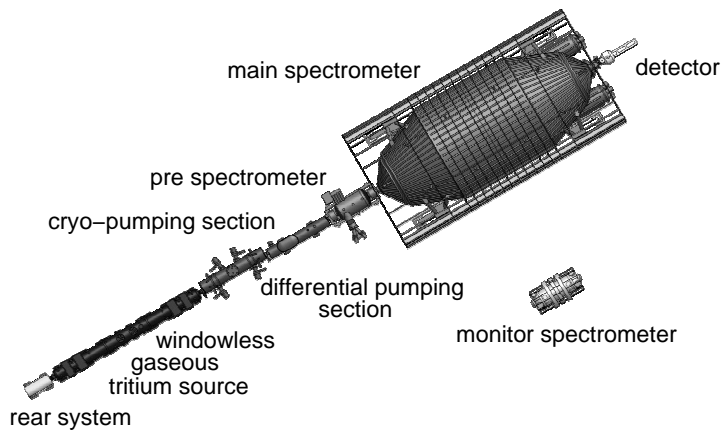}}
\caption{Schematic view of the 70~m long KATRIN experiment consisting of calibration and monitor rear system, 
windowless gaseous \ttwo -source, differential pumping and cryo-trapping section, small pre-spectrometer and large main spectrometer, 
segmented PIN-diode detector and separate monitor spectrometer.}\label{fig_katrin_setup}
\end{figure}

The international KATRIN collaboration has taken the challenge set 
by the astrophysics and particle physics arguments
to increase the sensitivity on the neutrino mass
down to 0.2~eV. It is currently setting up
an ultra-sensitive tritium \bdec\ experiment based on the successful 
MAC-E-Filter spectrometer technique and a very strong Windowless Gaseous Tritium Source (WGTS) 
\cite{KATRIN_loi,KATRIN_design_report} at the Karlsruhe Institute of Technology, Germany (formerly Forschungszentrum Karlsruhe).
Improving tritium $\beta$-spectroscopy by a factor of 100 in the
observable \mtwonue\ evidently requires  brute force, based on proven experimental concepts. 
It was decided, therefore, to build a spectrometer of MAC-E-Filter type with a diameter of 10 m, corresponding to a 100 times larger 
analyzing plane as compared to the pilot instruments at Mainz and Troitsk, feeded by \belec s from a high-luminosity windowless gaseous molecular 
tritium source.
Figure \ref{fig_katrin_setup} depicts a schematic plan  of the whole 70 m long setup. 

A decay rate of the order of $10^{11}$~Bq is aimed for in a source with a diameter of 9~cm requiring extraordinary demands in terms of size and cryo-techniques 
to handle the flux of $10^{19}$ \ttwo -molecules/s safely. \ttwo\
is injected at the midpoint of a 10~m long source tube kept at a temperature of 27~K by a 2-phase liquid neon bath. 
The integral column density of the source of $5 \cdot 10^{17}$~molecules/cm$^2$ has to be stabilized within 0.1~\%. 
Owing to background considerations, the \ttwo -flux entering the spectrometer should not exceed $10^5$ \ttwo -molecules/s. 
This will be achieved by differential pumping sections (DPS), followed by cryo-pumping sections (CPS) which trap residual \ttwo\ on argon frost at about 3~K. 
Each system reduces the throughput by $10^7$, which has been demonstrated for the cryo-pumping section by a dedicated experiment at Karlsruhe \cite{trap}. 
The \ttwo -gas collected by the DPS-pumps will be purified and recycled. 
All these components possess strong superconducting solenoids to guide the
\belec s from the source to the spectrometer 
within a magnetic flux tube of 191~T~cm$^2$. In summer 2009, the DPS arrived at Karlsruhe and is now being commissioned. 

\begin{figure}
\centerline{\includegraphics[angle=0,width=0.4\textwidth]{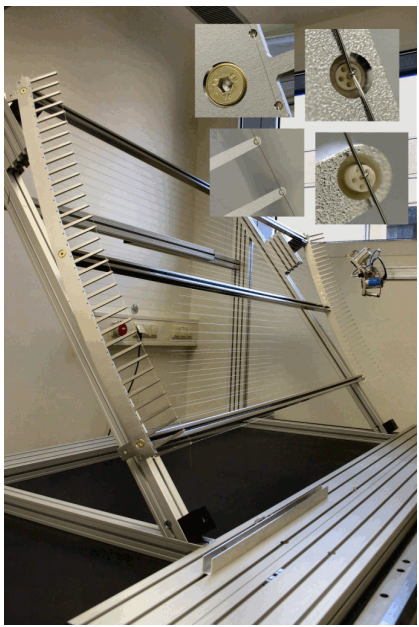}}
\caption{A double-layer wire electrode module on the 3-axis measurement table for quality assurance. 
The fixing of the wires inside the ceramics holders (see inserted smaller photos right) with the connecting wires is checked with a high-resolution camera,
whereas wire position and wire tension are monitored by a specially developed 2-dim. laser sensor.}
\label{fig:inner_electrode}
\end{figure}

A pre-spectrometer will transmit only the uppermost part of the \bspec\ into the main spectrometer in order to reduce the rate of background-producing 
ionization events therein. The entire pre- and main spectrometer vessels will each be put on their respective analyzing potentials, 
which are shifted inside by about -200~V, however, due to the installation of a background reducing inner screen grid system (fig. \ref{fig:inner_electrode}). 
A ratio of the maximum magnetic field in the pinch magnet over the minimum magnetic field in the central analyzing plane of the main spectrometer 
of 20000 provides an energy resolution of $\Delta E = 0.93$~eV near the tritium endpoint \ezero . 
The residual inhomogeneities of the electric retarding potential and the magnetic fields in the analyzing plane will be corrected by the 
spatial information from a 148 pixel PIN diode detector. 
Active and passive shields will minimize the background rate at the detector. 
Additional post-acceleration will reduce the background rate within the energy window of interest. 
Special care has to be taken to stabilize and to measure the retarding voltage. 
In addition to an ultra-precise HV divider \cite{thuemmler09}, 
the spectrometer of the former Mainz Neutrino Mass experiment will be operated at KATRIN as a high voltage monitor spectrometer 
which continuously measures the position of  the \kr -K32 conversion electron line at 17.8~keV, in parallel to the retarding energy of the main spectrometer. 
To that end its energy resolution has been refined to $\Delta E = 1$~eV.
 
The sensitivity limit of KATRIN on the neutrino mass 
has been simulated on the basis of a background rate of $10^{-2}$~cts/s, 
observed at Mainz and Troitsk under optimal conditions. Whether this small number can also be reached at the so much larger KATRIN instrument 
-- or even be lowered -- has yet to be proven. 
At Mainz the main residual background originated from secondary electrons from the walls/electrodes on high potential
caused by passing cosmic muons or by $\gamma$s from radioactive impurities.
Although there is a very effective magnetic shielding by the conservation of the magnetic flux, small
violation of the axial symmetry and other inhomogeneities allowed a fraction of about $10^{-5}$ of these secondary
electrons to reach the detector and to be counted as background.
After finishing tritium measurements in 2001, electrostatic  solutions were developed at Mainz, 
which strengthened shielding of surface electrons by an additional factor of $\approx 10$. 
This was achieved by covering the electrodes with negatively biased grids built from thin wires. 
Even more refined 2-layer wire electrode modules have been constructed for the KATRIN main spectrometer (fig. \ref{fig:inner_electrode})
to achieve a background suppression 
of 2 orders of magnitude. They are currently being installed.
The main spectrometer vessel has already 
passed successfully out-baking and out-gasing tests.

Since the KATRIN experiment will investigate only the very upper end of the \bspec , quite a few
systematic uncertainties will
be small because of excitation thresholds. Others systematics like the inelastic scattering fraction or the source intensity
will be controlled very precisely by measuring the column density online by an electron gun \cite{valerius09}, by 
keeping the temperature and pressure within the tritium source at the per mille level constant and by determining the tritium 
fraction of the gas in the source by laser Raman spectroscopy to the per mille level \cite{laserraman10}.
Therefore each systematic uncertainty contributes to the uncertainty of  \mtwonue\ with less than $0.0075$~eV$^2$, 
resulting in a total systematic uncertainty of 
$\Delta \mtwonue _{\rm sys} = 0.017$~eV$^2$. 
The total uncertainty will allow a sensitivity on \mnue\ of 0.2~eV to be reached. 
If no neutrino mass is observed, this sensitivity corresponds to an upper limit on \mnue\ of 0.2~eV at 90~\%~C.L, or, otherwise, to evidence for
(discovery of)  a non-zero neutrino mass value at $\mnue = 0.3$~eV (0.35~eV) with $3\sigma$ ($5 \sigma$) significance. 
For more details we refer to the KATRIN Design Report \cite{KATRIN_design_report}.

\subsection{``Source$=$detector'' configuration: $^{187}$Re $\beta$ decay experiments}

\rhenium\ is a second isotope suited to determine the neutrino mass. 
Due to the complicated electronic
structure of \rhenium\ and its long half life of $4.3 \cdot 10^{10}$~y 
the advantage of the 7 times lower
endpoint energy \ezero = 2.47~keV of \rhenium\ with respect to tritium can
only be exploited if the $\beta$ spectrometer measures  the entire
released energy, except that of the neutrino. This situation can
be realized by using a cryogenic bolometer as the $\beta$
spectrometer, which at the same time contains the $\beta$ emitter.

One disadvantage connected to the rhenium bolometer method is the fact that one
measures always the entire \bspec . Even for the case of the very
low endpoint energy of \rhenium , the relative fraction of \rhenium\ \bdec\ events
in the last eV below \ezero\ is of order $10^{-11}$ only (compare
to figure \ref{fig:beta_spec}). Considering the long time constant
of the signal of a cryogenic bolometer (typically several hundred
$\mu$s) pile-up is a severe problem, since it changes the spectral shape near
the endpoint. In order to limit the pile-up fraction to $\leq 10^{-4}$ 
only large arrays of cryogenic bolometers could deliver the signal rate needed.
Another technical challenge is the energy resolution of rhenium bolometers. Although cryogenic
bolometers with an energy resolution of a few eV have been produced
with other absorbers, this resolution has yet not been achieved with rhenium.

Two groups have started the field of \rhenium\ \bdec\ experiments
at Milan  (MiBeta) and Genoa (MANU):
The MANU experiment was using one metallic rhenium crystal of 1.5~mg working at a temperature of 100~mK and 
read out by Germanium doped thermistor. The $\beta$ environmental fine structure was observed for the first time
giving rise to a modulation of the shape of the \bspec\ by the interference of the out-going \belec\ wave with the
rhenium crystal \cite{manu_befs}. The spectrum near the endpoint allowed to set
an upper limit on the neutrino mass of $\mnue < 26$~eV. 
The MiBeta collaboration was using 10 crystals of AgReO$_4$ with a mass of about 0.25~mg each. The energy resolution
of a single bolometer was about 30~eV. One year of data taking resulted in an upper limit on the ``electron neutrino mass''
of $\mnue < 15$~eV \cite{mibeta}.

Both groups are now working together with other groups from different countries within the MARE collaboration \cite{mare}. 
The aim of MARE is to improve the energy resolution by new type of sensors (e.g. transition edge thermistors), to
increase the thermalization and read-out speed and to increase the number of detectors significantly. 
In MARE phase 1 300 detectors with an 
energy resolution of 20~eV and a read-out time of 100~$\mu$s should provide a sensitivity on the neutrino mass
of 2-3~eV. Although this is just the sensitivity which has already been reached in tritium \bdec\ experiments
this approach is interesting since it is very complementary in experimental techniques and systematic uncertainties.
Presently a first array of detectors is being commissioned with 30~eV energy resolution and with a read-out time of 250~$\mu$s.

The goal is, to increase in MARE phase 2 the number of detectors to 50000. Together with the energy resolution
goal of $\Delta E \leq 5$~eV and the aim to achieve a read-out and thermalization time of $\leq 5 \mu$s the MARE
could improve in phase 2 another order of magnitude in sensitivity on the neutrino mass. Of course this will only be
possible if the very challenging improvements in detector performance will really be reached and if no other problem 
with the bolometer technique (like any tiny process giving rise of an energy leakage, e.g. by soft photons or 
meta-stable excitations) will appear when improving the sensitivity on the observable \mtwo\ by 4 orders of magnitude.

\section{Conclusions}

The absolute neutrino mass scale is addressed by three different methods. The analysis of large scale structure data
and the anisotropies of the cosmic microwave background radiation are very sensitive but model dependent. The
search for neutrinoless double \bdec\ requires neutrinos to be their own antiparticles (Majorana neutrinos) is measuring
a coherent sum over all neutrino masses contributing to the electron neutrino with unknown phases.
Therefore the value of the neutrino mass cannot be determined very precisely, but
the discovery of neutrinoless double \bdec\ would be the detection of lepton flavor violation.
A few double \bdec\ experiments of the second generation are currently being commissioned and will deliver data
soon. Among the various ways to address the absolute neutrino mass scale
the investigation of the shape of \bdec\ spectra around the endpoint is the
only model-independent method. The KATRIN experiment is being setup at Karlsruhe and will start data taking in 2012, whereas
the MARE experiment is commissioning a small array of detectors starting MARE phase 1.
From both laboratory  approaches we expect in the coming years sensitivies on the neutrino mass  of {\cal O}(100)~meV.


\section*{Acknowledgments}
The work of the author is supported by the German Ministery for Education and Research BMBF.

\end{document}